\documentclass[a4paper,11pt]{article}%
\usepackage{amssymb,amsmath,amsfonts}
\usepackage{graphicx}
\usepackage{amsmath}
\usepackage{amsfonts}
\usepackage{amssymb}%
\providecommand{\U}[1]{\protect\rule{.1in}{.1in}}
\newtheorem{theorem}{Theorem}

\newtheorem{corollary}[theorem]{Corollary}

\newtheorem{idea memo}[theorem]{Idea Memo}

\newtheorem{remark}[theorem]{Remark}

\begin{document}

\title{Photon localization revisited}
\author{Izumi Ojima\thanks{E-mail: ojima@kurims.kyoto-u.ac.jp}\\Otsu 520-0105, Japan\\and \\Hayato Saigo\thanks{E-mail: h\_saigoh@nagahama-i-bio.ac.jp } \\Nagahama Institute of Bio-Science and Technology \\Nagahama 526-0829, Japan}
\date{}
\maketitle

\begin{abstract}
In the light of Newton-Wigner-Wightman theorem of localizability question, we
have proposed before a typical generation mechanism of effective mass for
photons to be localized in the form of polaritons owing to photon-media
interactions. In this paper, the general essence of this example model is
extracted in such a form as Quantum Field Ontology associated with
Eventualization Principle, which enables us to explain the mutual relations,
back and forth, between quantum fields and various forms of particles in the
localized form of the former.

\end{abstract}

\section{Introduction}

Extending the scope of our joint paper \cite{O-S} whose essense is summarized
in 1) and 2) below, we discuss in this paper the following points:

1) Starting from a specific problem of photon localization in the light of
Newton-Wigner-Wightman Theorem (Sec.2), we try here to clarify the
mathematical and conceptual relations among spatial points, localization
processes of physical systems into restricted regions in space (and time), in
contrast to the usual formulation dependent directly on the concepts of
particles and their masses (in a spacetime structure given in an \textit{a
priori} way). In this context, Wightman's mathematical formulation of the
Newton-Wigner paper plays an important role: On the basis of an imprimitivity
system on the 3-dimensional space, the absence of position observables is
shown to follow from the vanishing mass $m=0$ of a free photon.

2) We encounter here a sharp conflict between the mathematically clear-cut
negative result and the actual existence of experimental devices for detecting
photons in quantum optics which is impossible without the spatial localization
of detected photons. Fortunately, this conflict is resolved by the presence of
coupled modes of photons with material media which generates non-trivial
deviations of refractive index $n$ from $1$, or equivalently generates the
mass $m>0$, in such typical example cases as \textquotedblleft
polaritons\textquotedblright, as will be shown later (Sec.3.4).

3) Through the model example of polaritons, we learn that such fundamental
issues as related with mass and particles as its carriers should be viewed as
something variable dependent on the contexts and situations surrounding them.
Thus, we need and can elaborate on highly philosophical abstract questions
like \textquotedblleft what is a mass?\textquotedblright\ or \textquotedblleft
what are particles as mass points?\textquotedblright, in mathematically
accessible contexts. For this purpose, we certainly need to set up suitable
theoretical and/or mathematical frameworks and models so that they allow us to
systematically control the dynamics of our object systems coupled with their
external systems. Once this coupling scheme is established, the external
systems can be seen to serve as reference systems for the purpose of
describing the object systems and the processes carried out by them. Such a
framework and methodology are available in the form of the Tomita's integral
decomposition theorem (Sec. 4.3) viewed from the standpoint of
\textquotedblleft quadrality scheme\textquotedblright\ based on
\textquotedblleft Micro-Macro duality\textquotedblright\ (Sec. 3.2 \& Sec. 4.2).

4) For instance, the delicate choice between 4-dimensional spacetime and
3-dimensional spatial setting up involved in Wightman's theorem can be
naturally understood as the choice of pertinent variables to a given context.
In the light of Tomita's theorem this issue is seen in such a form as the
choice between central vs. subcentral decomposition measures of a relevant
state. A satisfactory understanding of fundamental concepts of space (and
time) coordinates and velocities is attainable in the scheme and, at the same
time, crucial premise underlying such comprehension is the understanding that
these concepts are never among pre-existing attributes inherent in the object
system but are epigenetic properties emerging through what is to be called the
\textquotedblleft eventualization processes\textquotedblright\ as will be
explained in Sec.5. These epigenetic aspects are closely related with the
choices of different contexts of placing an object system and the boundary
conditions specifying various different choices of subalgebras of central
observables, reflected in the choices of subcentral (or central) measures
appearing in Tomita's theorem (Sec. 4.3).

5) While the above explanation guarantees the naturality and genericity of the
polariton picture mentioned in 2), as one of the typical explicit examples for
making photons localizable, the freedom in choices of subcentral measures
clarifies their speciality in the \textit{spatial homogeneity} of mass
generation. In fact, under such conditions that the spatial homogeneity is
\textit{not} required, many such forms of photon localizations are allowed as
Debye shielding, various forms of dressed photons, among which cavity QED can
equally be understandable.

6) Along this line of thought, it becomes also possible to compare and unify
various other forms of localizations and of their \textquotedblleft
leakages\textquotedblright\ at the same time:\ For instance, the presence of
non-vanishing mass $m$ can be viewed as an index of timelike and
spacetime-homogeneous parameter of leakage from spatial localization as
exhibited by the decay rate $\propto e^{-mr}$ of correlation functions in
clustering limit. On the other hand, the decay width $\Gamma$ in the energy
spectrum can also be interpreted as a time-homogeneous parameter of leakage
from chronological localizations of resonance modes (as exhibited through the
decay rate $\propto e^{-\Gamma/2}$ of relaxation of correlations). (To be
precise, it is more appropriate to regard the inverse of $m$ and $\Gamma$ as
leakages.) The tunneling rate $\propto\sqrt{\left\vert E-V\right\vert }$ can
be interpreted as the leakage rate of spatial localization materialized by the
potential barrier $V$. 

7) The universality, naturality and the necessity of the present standpoint is
verified by the above considerations in terms of subcentral measures and of
the corresponding commutative algebras $\mathcal{B}$. On the basis of the
bidirectionality between quantum fields and particles, moreover, such a
unified viewpoint will be meaningful that the microscopic quantum systems
consisting of quantum fields can be controlled and designed from the macro
side via the control of quantum fields.

8) To make sure of the above possibility, it would be important to recognize
the constitution of the macroscopic levels in close relations with the
microscopic quantum regimes. This question is answered in terms of the word,
\textquotedblleft eventualization processes\textquotedblright, which can be
mathematically described as the filtered \textquotedblleft
cones\textquotedblright\ to amplify the connections between Macro and Micro
(which is analogous to the forcing method in the context of foudations of
mathematics), with Micro ends given by the dynamics of quantum fields and
Macro ones by the pointlike events as the apices of cones of eventualizations.

\section{Newton-Wigner-Wightman Theorem}

In 1949, Newton and Wigner \cite{N-W} raised the question of localizability of
single free particles. They attempted to formulate the properties of the
localized states on the basis of natural requirements of relativistic covariance.

\paragraph{}

Physical quantities available in this formulation admitting direct physical
meaning are restricted inevitably to the generators of Poincar\'{e} group
$\mathcal{P}_{+}^{\uparrow}=\mathbb{R}^{4}\rtimes L_{+}^{\uparrow}$ (with
$L_{+}^{\uparrow}$ the orthochronous proper Lorentz group) which is locally
isomorphic to the semi-direct product $\mathcal{H}_{2}(\mathbb{C})\rtimes
SL(2,\mathbb{C})$ of the Jordan algebra $\mathcal{H}_{2}(\mathbb{C})$ of
hermitian $(2\times2)$-matrices and $SL(2,\mathbb{C})$, consisting of the
energy-momentum vector $P_{\mu}$ and of the Lorentz generators $M_{\mu\nu}$
(composed of angular momenta $M_{ij}$ and of Lorentz boosts $M_{0i}$). The
problem is then to find conditions under which \textquotedblleft position
operators\textquotedblright\ can naturally be derived from the Poincar\'{e}
generators $(P_{\mu},M_{\mu\nu})$. In \cite{N-W}, position operators have been
shown to exist in massive cases in an essentially unique way for
\textquotedblleft elementary\textquotedblright\ systems in the sense of the
irreducibility of the corresponding representations of $\mathcal{P}%
_{+}^{\uparrow}$ so that localizability of a state can be defined in terms of
such position operators. In massless cases, however, no localized states are
found to exist in the above sense. That was the beginning of the story.

Wightman \cite{WIG} clarified the situation by recapturing the concept of
\textquotedblleft localization\textquotedblright\ in quite a general form as
follows. In place of the usual approaches with unbounded generators of
position operators, he has formulated the problem in terms of their spectral
resolution in the form of axioms (i)-(iii) :

\begin{description}
\item[(i)] The spectral resolution of position operators: It is defined by a
family $\mathcal{B}(\mathbb{R}^{3})\ni\Delta\longmapsto E(\Delta)\in
Proj(\mathfrak{H})$ of projection-valued measures $E(\Delta)$ in a Hilbert
space $\mathfrak{H}$ defined for each Borel subset $\Delta$ of $\mathbb{R}%
^{3}$, characterized by the following properties (ia), (ib), (ic):

\begin{description}
\item[(ia)] $E(\Delta_{1}\cap\Delta_{2})=E(\Delta_{1})E(\Delta_{2})$;

\item[(ib)] $E(\Delta_{1}\cup\Delta_{2})=E(\Delta_{1})+E(\Delta_{2})$, if
$\Delta_{1}\cap\Delta_{2}=\phi$;

\item[(ic)] $E(\mathbb{R}^{3})=1$;
\end{description}

\item[(ii)] Physical interpretation of $E(\Delta)$: When the system is
prepared in a state $\omega$, the expectation value $\omega(E(\Delta))$ of a
spectral measure $E(\Delta)$ gives the probability for the system to be found
in a localized region $\Delta$;

\item[(iii)] Covariance of the spectral resolution: Under a transformation
$(\mathbf{a},\mathcal{R})$\newline with a spatial rotation $\mathcal{R}$
followed by a spatial translation $\mathbf{a}$, a Borel subset $\Delta$ is
transformed into $\mathcal{R}\Delta+\mathbf{a}$. The corresponding unitary
implementer is given in $\mathfrak{H}$ by $U(\mathbf{a},\mathcal{R})$, which
represents $(\mathbf{a},\mathcal{R})$ covariantly on $E$ in such a way that
\[
E(\Delta)\rightarrow E(\mathcal{R}\Delta+\mathbf{a})=U(\mathbf{a}%
,\mathcal{R})E(\Delta)U(\mathbf{a},\mathcal{R})^{-1}.
\]

\end{description}

Note that, in spite of the relevance of the relativistic covariance,
localizability discussed above is the \textit{localization of states in space
at a given time }formulated in terms of \textit{spatial} translations
$\mathbf{a}$ and rotations $\mathcal{R}$, respectively. To understand the
reason, one should imagine the situation with the axioms (i)-(iii) replaced
with those for the whole spacetime; then the CCR relations hold between
4-momenta $p_{\mu}$ and space-time coordinates $x^{\nu}$, which implies the
Lebesgue spectrum covering the whole $\mathbb{R}^{4}$ for both observables
$\hat{p}_{\mu}$ and $\hat{x}^{\nu}$. Therefore any such physical requirements
as the spectrum condition or as the mass spectrum cannot be imposed on the
energy-momentum spectrum $\hat{p}_{\mu}$, and hence, the concept of
localizability in space-time does not make sense.

According to Mackey's theory of induced representations, Wightman's
formulation can easily be seen as the condition for the family of operators
$\left\{  E(\Delta)\right\}  $ to constitute a \textit{system of
imprimitivity} (\cite{MAC}) under the action of the unitary representation
$U(\mathbf{a},\mathcal{R})$ in $\mathfrak{H}$ of the three-dimensional
Euclidean group $SE(3):=\mathbb{R}^{3}\rtimes SO(3)$ given by the semi-direct
product of the spatial translations $\mathbb{R}^{3}$ and the rotation group
$SO(3)$. In a more algebraic form, the pair $(E,U)$ can also be viewed as a
\textit{covariant W*-dynamical system} $L^{\infty}(\mathbb{R}^{3}%
)\underset{\tau}{\curvearrowleft}SE(3),[\tau_{(\mathbf{a},\mathcal{R)}%
}(f)](\mathbf{x}):=f(\mathcal{R}^{-1}(\mathbf{x}-\mathbf{a}))$, given by the
covariant *-representation $E:L^{\infty}(\mathbb{R}^{3})\ni f\longmapsto
E(f)=\int f(\mathbf{x})dE(\mathbf{x})\in B(\mathfrak{H})$, s.t. $E(\chi
_{\Delta})=E(\Delta)$, of the commutative algebra $L^{\infty}(\mathbb{R}^{3})$
generated by the position operators acted on by $SE(3)$ characterized by the
\textit{covariance condition}:
\begin{align}
E(\tau_{(\mathbf{a},\mathcal{R)}}(f))  &  =U(a,\mathcal{R})E(f)U(a,\mathcal{R}%
)^{-1}\label{imprim}\\
\text{ \ \ \ for }f  &  \in L^{\infty}(\mathbb{R}^{3}),(\mathbf{a}%
,\mathcal{R)\in}SE(3).\nonumber
\end{align}
As will be seen later, this algebraic reformulation turns out to be useful for
constructing coupled systems of photon degrees of freedom with matter systems,
which play the crucial roles in observing or measuring the former in the
actual situations. Thus Wightman's formulation of the Newton-Wigner
localizability problem is just to examine whether the Hilbert space
$\mathfrak{H}$ of the representation $(U,\mathfrak{H})$ of $SE(3)$ can
accommodate a representation $E$ of the algebra $L^{\infty}(\mathbb{R}^{3})$
consisting of position operators, covariant under the action of $SE(3)$ in the
sense of Eq. (\ref{imprim}).

Applying Mackey's general theory to the case of three-dimensional Euclidean
group $SE(3)$, Wightman proved the following fundamental result as a purely
kinematical consequence:

\begin{theorem}
[\cite{WIG}, excerpt from theorem 6 and 7]A Lorentz covariant massive system
is always localizable. The only localizable massless elementary system (i.e.
irreducible representation) has spin zero.
\end{theorem}

\begin{corollary}
A free photon is not localizable.
\end{corollary}

The essential mechanism causing (non-)localizability in the sense of
Newton-Wigner-Wightman can be found in the structure of Wigner's little
groups, the stabilizer groups of standard 4-momenta on each type of
$\mathcal{P}_{+}^{\uparrow}$-orbits in $p$-space.

When $m\neq0$, the little group corresponding to the residual degrees of
freedom in a rest frame is the group $SO(3)$ of spatial rotations. As a
consequence, \textquotedblleft the space of rest frames\textquotedblright%
\ becomes $SO(1,3)/SO(3)\cong\mathbb{R}^{3}$. The physical meaning of this
homeomorphism is just a correspondence between a rest frame $r\in
SO(1,3)/SO(3)$ for registering positions\ and a boost $k\in SO(1,3)$ required
for transforming a fixed rest frame $r_{0}$ to the chosen one $r=kr_{0}$. The
universality (or, independence for the choice the frame) of positions\ is
recovered up to Compton wavelength $h/(mc)$, again due to massiveness.

\begin{remark}
Here the coordinates of rest frames\ just plays the role of the order
parameters (or, \textquotedblleft sector parameters\textquotedblright) on each
$\mathcal{P}_{+}^{\uparrow}$-orbit as the space of \textquotedblleft
degenerate vacua\textquotedblright\ associated with certain of symmetry
breaking, which should play the roles of position operators appearing in the
imprimitivity system.
\end{remark}

In sharp contrast, there is \textit{no} rest frame for a massless particle:
Its little group is isomorphic to the two-dimensional Euclidean group
$SE(2)=\mathbb{R}^{2}\rtimes SO(2)$ (locally isomorphic to $\mathbb{C}\rtimes
U(1)$), whose rotational generator corresponds to the helicity. Since the
other two translation generators corresponding to gauge transformations span
\textit{non-compact} directions in distinction from the massive cases with a
compact $SO(3)$, the allowed representation (without indefinite inner product)
is only the trivial one which leaves the transverse modes invariant, and
hence, the little group cannot provide position operators in the massless case.

After the papers by Newton and Wigner and by Wightman, many discussions have
been developed around the photon localization problem. As far as we know, the
arguments seem to be divided into two opposite directions, one relying on
purely dynamical bases \cite{HAA} and another on pure kinematics \cite{FLA},
where it is almost impossible to find any meaningful agreements. Below we
propose an alternative strategy based on the concept of \textquotedblleft
effective mass\textquotedblright, which can provide a reasonable
reconciliation between these conflicting ideas because of its
\textquotedblleft kinematical\textquotedblright\ nature arising from some
dynamical origin.

\section{Polariton as a Typical Model of Effective Mass Generation}

\subsection{Physical roles played by coupled external system}

In spite of the above theoretical difficulty in the localizability of photons,
however, it is a plain fact that almost no experiments can be performed in
quantum optics where photons must be registered by \textit{localized}
detectors. To elaborate on this problem, we will see that it is indispensable
to reexamine the behaviour of a photon in composite systems coupled with some
external system such as material media constituting apparatus without which
any kind of measurement processes cannot make sense. For this purpose, the
above group-theoretical analysis of localizability of kinematical nature
should be extended to incorporate algebraic aspects involved in the formation
of a coupled dynamics between photons to be detected and the measuring devices
consisting of matters.

Our scheme of the localization for photons can be summarized as follows:

\begin{itemize}
\item Photons are coupled with external system into a composite system with a
coupled dynamics.

\item Positive effective mass emerges in the composite system.

\item Once a positive effective mass appears, Wightman's theorem itself
provides the \textquotedblleft kinematical basis\textquotedblright\ for the
localization of a photon.
\end{itemize}

From our point of view, therefore, this theorem of Wightman's interpreted
traditionally as a no-go theorem against the localizability becomes actually
an affirmative support for it. It conveys such a strongly selective meaning
(which will be discussed in detail in Sec.4) that, whenever a photon is
localized, it should carry\ a non-zero effective mass.

In the next subsection, we explain the meaning of our scheme from a physical
point of view.

\subsection{How to define effective mass of a photon}

As a typical example of our scheme, we focus first on a photon interacting
with homogeneous medium, in the case of the monochromatic light with angular
frequency $\omega$ as a classical light wave. For simplicity, we neglect here
the effect of absorption, that is, the imaginary part of refractive index.
When a photon interacting with matter can be treated as a single particle, it
is natural to identify its velocity $\mathbf{v}$ with the \textquotedblleft
signal velocity\textquotedblright\ of light in medium. The relativistic total
energy $E$ of the particle should be related to $v:=\sqrt{\mathbf{v\cdot v}}$
by its mass $m_{\mathsf{eff}}$:
\begin{equation}
E=\frac{m_{\mathsf{eff}}c^{2}}{\sqrt{1-\frac{v^{2}}{c^{2}}}} \label{energy}%
\end{equation}
Since $v$ is well known to be smaller than the light velocity $c$
(theoretically or experimentally), $m_{\mathsf{eff}}$ is positive (when the
particle picture above is valid). Then we may consider $m_{\mathsf{eff}}$ as
the relativistic \textquotedblleft effective (rest) mass of a
photon\textquotedblright, and identify its momentum $\mathbf{p}$ with
\begin{equation}
\text{$\mathbf{p}$}=\frac{m_{\mathsf{eff}}\text{$\mathbf{v}$}}{\sqrt
{1-\frac{v^{2}}{c^{2}}}}. \label{momentum1}%
\end{equation}
Hence, as long as \textquotedblleft an interacting photon\textquotedblright%
\ can be well approximated by a single particle, it should be massive,
according to which its \textquotedblleft localization
problem\textquotedblright\ is resolved. The validity of this picture will be
confirmed later in the next subsection.

The concrete forms of energy/momentum are related to the Abraham-Minkowski
controversy \cite{ABR, MIN, BAR} and modified versions of Einstein/de Broglie
formulae \cite{O-S}.

Our argument itself, however, does not depend on the energy/momentum formulae.
The only essential point is that a massless particle can be made massive
through some interactions. That is, while a free photon satisfies
\begin{equation}
E_{\mathrm{free}}^{2}-c^{2}p_{\mathrm{free}}^{2}=0,
\end{equation}
an interacting photon satisfies
\begin{equation}
E^{2}-c^{2}p^{2}=m_{\mathsf{eff}}^{2}c^{4}>0.
\end{equation}
To sum up, an \textquotedblleft interacting photon\textquotedblright\ can gain
a positive effective mass, while a \textquotedblleft free
photon\textquotedblright\ remains massless! This is the key we have sought
for. We note, however, the present argument is based on the assumption that
\textquotedblleft a photon dressed with interactions\textquotedblright\ can be
viewed as a single particle.
We proceed to consolidate the validity of this picture, especially the
existence of particles whose effective mass is produced by the interactions,
analogous to Higgs mechanism: Such a universal model for photon localization
certainly exists, which is based on the concept of polariton, well known in
optical and solid physics.

\subsection{Polariton picture}

In these areas of physics, the propagation of light in a medium is viewed as
follows: By the interaction between light and matter, creation of an
\textquotedblleft excition (an excited state of polarization field above the
Fermi surface)\textquotedblright\ and annihilation of a photon will be
followed by annihilation of an exciton and creation of a photon, $\cdots$, and
so on. This chain of processes itself is often considered as the motion of
particles called \textit{polaritons} (in this case \textquotedblleft
exciton-polaritons\textquotedblright), which constitute particles associated
with the coupled wave of the polarization wave and electromagnetic wave.


The concept of polariton has been introduced to develop a microscopic theory
of electromagnetic interactions in materials (\cite{FAN}, \cite{HOP}).
Injected photons become polaritons by the interaction with matter. As
exiton-phonon interaction is dissipative, the polariton picture gives a
scenario of absorption. It has provided an approximation better than the
scenarios without it. Moreover, the group velocity of polaritons discussed
below gives another confirmation of the presence of an effective mass.

As is well known, permittivity $\epsilon(\omega)$ is given by the following
equality,%
\begin{equation}
\epsilon(\omega)=n^{2}=\frac{c^{2}k^{2}}{\omega^{2}},
\end{equation}
and hence, we can determine the dispersion relation (between frequency and
wave number) of polariton once the formula of permittivity is specified. In
general, this dispersion relation implies branching, analogous to the Higgs
mechanism. The signal pulse correponding to each branch can also be detected
in many experiments, for example, in \cite{MAS} cited below.

In the simple case, the permittivity is given by the transverse frequency
$\omega_{T}$ of exciton's (lattice vibration) as follows:
\begin{equation}
\epsilon(\omega)=\epsilon_{\infty}+\frac{\omega_{T}^{2}(\epsilon_{st}%
-\epsilon_{\infty})}{\omega_{T}^{2}-\omega^{2}},
\end{equation}
where $\epsilon_{\infty}$ denotes $\lim_{\omega\rightarrow\infty}%
\epsilon(\omega)$ and $\epsilon_{st}=\epsilon(0)$ (static permittivity). With
a slight improvement through the wavenumber dependence of the exciton energy,
the theoretical result of polariton group velocity $\frac{\partial\omega
}{\partial\mathbf{k}}<c$ based on the above dispersion relation can explain
satisfactorily experimental data of the passing time of light in materials
(for example, \cite{MAS}). This strongly supports the validity of the
polariton picture.

From the above arguments, polaritons can be considered as a universal model of
the \textquotedblleft interacting photons in a medium\textquotedblright\ in
the previous section. The positive mass of a polariton gives a solution to its
\textquotedblleft localization problem\textquotedblright. Conversely, as the
\textquotedblleft consequence\textquotedblright\ of Wightman's theorem, it
follows that \textquotedblleft all\textquotedblright\ physically accessible
photons as particles which can be localized are more or less polaritons (or
similar particles) because only the interaction can give a photon its
effective mass, if it does not violate particle picture.

\section{Effective Mass Generation in General}

\subsection{Toward general situations}

In the last subsection we have discussed that the interaction of photons with
media can cause their localization by giving effective masses to them. Then a
natural question arises: Is the exsistence of media a necessary condition for
the emergence of effectve photon mass? The answer is no: In fact, light beams
with finite transvese size have group velocities less than $c$.

In a recent publication \cite{Gio}, Giovannni et al., show experimentally that
even in vacuum photons (in the optical regime) travel at the speed less than
$c$ when it is transversally structured, such as Bessel beams or Gaussian
beams, by measuring a change in the arrival time of time-correlated photon
pairs. They show a reduction in the velocity of photons in both a Bessel beam
and a focused Gaussian one. Their work highlights that, even in free space,
the invariance of the speed of light only applies to plane waves, i.e., free photons.

From our viewpoint, this result can be understood quite naturally in the light
of the Newton-Wigner-Wightman theorem. As we have seen, the theorem states
that every localizable elementary system (particle) with spin must be massive.
It implies that photons in the real world should travel less than $c$, in any
conditions, which makes the probability distribution of its position
well-defined without contradicting with the presence of spin. Hence,
transversally structured photons should become slow.

The scenario also applies to more general settings. Any kinds of boundary
conditions with finite volume (like cavity), or even nanoparticles in the
context of dressed photons \cite{Ohtsu}, will make photons heavier and slower,
even without medium!

\subsection{Wightman's theorem re-interpreted as the \textquotedblleft
basis\textquotedblright\ for localization}

Our general scheme of the localization for photons can be depicted as follows,
whose essense can be understood in accordance with the basic formulation of
\textquotedblleft quadrality scheme\textquotedblright\ \cite{OJI2} underlying
the Micro-Macro duality \cite{Unif03, OJI1}:
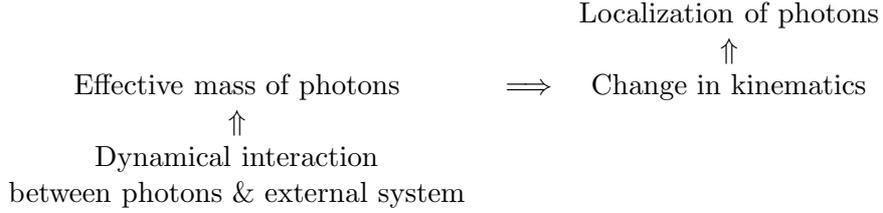
\[%
\begin{array}
[c]{ccc}
&  & \text{Localization of photons}\\
&  & \Uparrow\\
\text{Effective mass of photons} & \Longrightarrow & \text{Change in
kinematics}\\
\Uparrow &  & \\%
\begin{array}
[c]{c}%
\text{Dynamical interaction}\\
\text{between photons \& external system}%
\end{array}
&  &
\end{array}
\]
\newline In order to actualize the physical properties of a given system such
as photons driven by an invisible microscopic dynamics, it is necessary for it
to be coupled with some external measuring system through which a composite
system is formed. According to this formation of coupled dynamics, the
kinematics controlling the observed photons are modified and what can be
actually observed is a result of this changed kinematics, realized in our case
in the form of localized photons.

\subsection{Tomita's theorem of integral decomposition of a state}

Before going into the details of mass generation mechanisms, we examine here
the theoretical framework relevant to our context. From the mathematical
viewpoint, an idealized form of constructing a coupled system of the object
system with an external reference one can be found conveniently in Tomita's
theorem of integral decomposition of a state as follows:

\begin{theorem}
[Tomita \cite{BR1}]For a state $\omega$ of a unital C*-algebra $\mathcal{A}$,
the following three sets are in a 1-to-1 correspondence:

\begin{enumerate}
\item subcentral measures $\mu$ (pseudo-)supported by the space
$F_{\mathcal{A}}$ of factor states on $\mathcal{A}$;

\item abelian von Neumann subalgebras $\mathcal{B}$ of the centre
$\mathfrak{Z}_{\pi_{\omega}}(\mathcal{A})=\pi_{\omega}(\mathcal{A}%
)^{\prime\prime}\cap\pi_{\omega}(\mathcal{A})^{\prime}$;

\item central projections $C$ on $\mathfrak{H}_{\omega}$ such that%
\begin{equation}
C\Omega_{\omega}=\Omega_{\omega},\text{ \ \ \ }C\pi_{\omega}(\mathcal{A}%
)C\subset\{C\pi_{\omega}(\mathcal{A})C\}^{\prime}.
\end{equation}

\end{enumerate}

If $\mu$, $\mathcal{B}$ and $C$ are in the above correspondence, then the
following relations hold:

\begin{enumerate}
\item[(i)] $\mathcal{B}=\{\pi_{\omega}(\mathcal{A})\cup\{C\}\}^{\prime}$;

\item[(ii)] $C=[\mathcal{B}\Omega_{\omega}]$: projection operator onto the
subspace spanned by $\mathcal{B}\Omega_{\omega}$;

\item[(iii)] $\mu(\hat{A}_{1}\hat{A}_{2}\cdots\hat{A}_{n})=\langle
\Omega_{\omega}|\ \pi_{\omega}(A_{1})C\ \pi_{\omega}(A_{2})C\cdots
C\pi_{\omega}(A_{n})\Omega_{\omega}\rangle$ \newline for $A_{1},A_{2}%
,\cdots,A_{n}\in\mathcal{A}$;

\item[(iv)] The map $\kappa_{\mu}:L^{\infty}(E_{\mathcal{A}},\mu
)\rightarrow\mathcal{B}$ defined by
\begin{equation}
\langle\Omega_{\omega}|\ \kappa_{\mu}(f)\pi_{\omega}(A)\Omega_{\omega}%
\rangle=\int d\mu(\omega^{\prime})f(\omega^{\prime})\omega^{\prime}(A)
\end{equation}
for $f\in L^{\infty}(E_{\mathcal{A}},\mu)$ and $A\in\mathcal{A}$ is a
*-isomorphism, satisfying the following equality for $A,B\in\mathcal{A}$:
\begin{equation}
\kappa_{\mu}(\hat{A})\pi_{\omega}(B)\Omega_{\omega}=\pi_{\omega}%
(B)C\pi_{\omega}(A)\Omega_{\omega}.
\end{equation}

\end{enumerate}
\end{theorem}

Some vocabulary in the above need be explained: The space $F_{\mathcal{A}}$ of
factor states on $\mathcal{A}$ is the set of all the factor states $\varphi$
whose (GNS) representations $\pi_{\varphi}$ have trivial centres:
$\pi_{\varphi}(\mathcal{A})^{\prime\prime}\cap\pi_{\varphi}(\mathcal{A}%
)^{\prime}=\mathbb{C}1_{\mathfrak{H}_{\varphi}}$. This $F_{\mathcal{A}}$
divided by the quasi-equivalence relation $\approx$ defined by the unitary
equivalence up to multiplicity, $F_{\mathcal{A}}/\approx$ plays the role of
sector-classifying space (or, sector space, for short) whose elements we call
\textquotedblleft sectors\textquotedblright\ mathematically or
\textquotedblleft pure phases\textquotedblright\ physically. Then Tomita's
theorem plays a crucial role in verifying mathematically the so-called Born
rule \cite{OOS13} postulated in quantum theory in physics.

Via the definition $\widehat{A}(\rho):=\rho(A)$, $\rho\in E_{\mathcal{A}}$,
any element $A\in\mathcal{A}$ can be expressed by a continuous function
$\widehat{A}:E_{\mathcal{A}}\rightarrow\mathbb{C}$ on the state space
$E_{\mathcal{A}}$. Among measures on $E_{\mathcal{A}}$, a measure $\mu$ is
called \textit{barycentric} for a state $\omega\in E_{\mathcal{A}}$ if it
satisfies $\omega=\int_{E_{\mathcal{A}}}\rho d\mu(\rho)\in E_{\mathcal{A}}$
and is said to be \textit{subcentral} if linear functionals $\int_{\Delta}\rho
d\mu(\rho)$ and $\int_{E_{\mathcal{A}}\backslash\Delta}\sigma d\mu(\sigma)$ on
$\mathcal{A}$ are disjoint for any Borel set $\Delta\subset E_{\mathcal{A}}$,
having no non-vanishing intertwiners between them: i.e., $T\int_{\Delta}%
\pi_{\rho}(A)d\mu(\rho)=\int_{E_{\mathcal{A}}\backslash\Delta}\pi_{\sigma
}(A)d\mu(\sigma)T$ for $\forall A\in\mathcal{A}$ implies $T=0$. If the abelian
subalgebra $\mathcal{B}$ in the above theorem is equal to the centre
$\mathcal{B}=\mathfrak{Z}_{\pi_{\omega}}(\mathcal{A})$, the measure $\mu$ is
called the central measure of $\omega$, determined uniquely by the state
$\omega$ and the corresponding barycentric decomposition $\omega
=\int_{F_{\mathcal{A}}}\rho d\mu(\rho)$ is called the central decomposition of
$\omega$. This last concept plays crucial roles in establishing precisely the
bi-directional relations between microscopic and macroscopic aspects in
quantum theory, as has been exhibited by the examples of \textquotedblleft
Micro-Macro duality\textquotedblright\ (see, for instance, \cite{Unif03, OJI1}).

At first sight, the distinction between central and subcentral may look too
subtle, but it plays important roles in different treatments, for instance,
between \textit{spatial} and \textit{spacetime} degrees of freedom in
Wightman's theorem concerning the localizability, as mentioned already after
the theorem. In this connection, we consider the problem as to how classically
visible configurations of electromagnetic field can be specified in close
relation with its microscopic quantum behaviour, for the purpose of which most
convenient concept seems to be the coherent state and the Segal-Bargmann
transform associated with it. Since coherent states are usually treated within
the framework of quantum mechancs for systems with the \textit{finite degrees
of freedom}, the aspect commonly discussed is the so-called
\textit{overcompleteness relations }due to the non-orthogonality,
$\langle\alpha|\beta\rangle\neq0$, between coherent states $\hat{a}%
|\alpha\rangle=\alpha|\alpha\rangle$ with different coherence parameters
$\alpha\neq\beta$.

We note that, in connection with Tomita's theorem, a composite system arises
in such a form as $\mathcal{A}\otimes C(\Sigma)$ consisting of the object
system $\mathcal{A}$ and of the external system $\Sigma(\subset F_{\mathcal{A}%
})$ to which measured data are to be registered through measurement processes
involving $\mathcal{A}$. In this scheme, the universal reference system
$\Sigma$ can be viewed naturally \textit{emergent} from the object system
$\mathcal{A}$ itself just as the classifying space of its sector structure.
Then, via the \textit{logical extension} \cite{OjOz92} to parametrize the
object system $\mathcal{A}$ by its sectors in $\Sigma$, an abstract model of
\textit{quantum fields} $\varphi:\Sigma\rightarrow\mathcal{A}$ can be created,
constituting a crossed product $\varphi\in\mathcal{A}\rtimes\widehat
{\mathcal{U}(\Sigma)}$ (via the co-action of the structure group
$\mathcal{U}(\Sigma)$ of $\Sigma$). Thus, the above non-orthogonality can be
resolved by the effects of the classifying parameters of sectors $\Sigma$ in
$F_{\mathcal{A}}$. As a result, we arrive at the quantum-probabilistic
realization of coherent states in such a form as the \textquotedblleft
exponential vectors\textquotedblright\ treated by Obata \cite{Obata} in the
context of \textquotedblleft Fock expansions\textquotedblright\ of white
noises. What is important conceptually in this framework is the
\textit{analyticity} due to the Segal-Bargmann transform and the associated
\textit{reproducing kernel} (RS) to be identified through the projection
operator $P$ in $L^{2}(\Sigma,d\mu)$ onto its subspace $\mathcal{H}%
L^{2}(\Sigma,d\mu)$ of coherent states expressed by holomorphic functions on
$\Sigma$ \cite{Hall}, where $d\mu$ denotes the Gaussian measure.

As commented briefly above, we can find various useful relations and
connections of quantum theory in terms of the concept of \textquotedblleft
quantum fields\textquotedblright. From this viewpoint, we elaborate on its
roles in attaining a transparent understanding of the mutual relations among
fields, particles and mass in the next section.

\section{Quantum Field Ontology}

\subsection{From particles to fields}

As we have discussed in Sec.4, the effective mass generating scenario applies
to general settings. Any kinds of boundary conditions with finite volume (like
cavity) will make photons heavier and slower, even without medium. This fact
itself leads to a paradoxical physical question --- how can the boundary
condition affect a particle traveling in vacua? What is a spooky action
through vacua?

Our answer is quite simple: In fact a photon is not \textquotedblleft a
particle traveling in vacua\textquotedblright. It is just a field filling the
space time, before it \textquotedblleft becomes\textquotedblright\ a particle,
or more rigorously, before it appears in a particle-like event caused via the
interaction (energy-momentum exchange with external system). As we will
discuss in this section, it is quite unreasonable to imagine a photon as a
traveling particle unless any kinds of interaction is there.

Based on the arguments above, we discuss the limitation of particle concept in
connection with a new physical interpretation of Newton-Wigner-Wightman analysis.

To begin with, we should mention that this concept involves a strong
inconsistency with particle concept which seems to have been forgotten at some
stage in history. In fact, the concept of a classical massless point particle
with non-zero spin cannot survive special relativity with the worldline of
such a particle obscured by the spin: Instead of being a purely
\textquotedblleft internal\textquotedblright\ degree of freedom, the spin
causes kinematical extensivity of the particle which is exhibited in a boost
transformation, as is pointed out by Bacry in \cite{Bac}.

The result of Newton-Wigner-Wightman analysis can be understood to show that
this inconsistency cannot be eliminated by generalizing the problem in the
context of quantum theory: A massless particle cannot be localized unless the
spin is zero. Even in the massive case, the concept of localization is not
independent of the choice of reference frames. There is no well-defined
concept of \textquotedblleft spacetime localization\textquotedblright\ as we
have mentioned.

These facts are consistent with the idea that the position is not a clear cut
a priori concept but an emergent property. Instead of a point particle,
therefore, we should find something else\ having spacetime structure to
accommodate events in point-like forms, which is nothing but the quantum
field. In other words, the Newton-Wigner-Wightman analysis should be
re-interpreted as \textquotedblleft the existence proof of a quantum
field\textquotedblright, showing its inevitability.

\subsection{From fields to particles: Principle of eventualization}

This does not mean that particle-like property is artificial nor fictional. On
the contrary, point-like events do take place in any kind of elementary
processes of quantum measurement such as exposure on a film, photon counting,
and so on.

This apparent contradiction is solved if we adopt the universality of the
indeterminate processes emerging point-like events (energy-momentum exchanges)
from quantum fields via formation of composite system with external systems
(like media or systems giving boundary conditions), even the latter coming
from the part of the degrees of freedom of quantum fields. Let us call these
fundamental processes as \textit{eventualization}. From our viewpoint, the
most radical implication of Newton-Wigner-Wightman analysis is that we should
abandon the ontology based on na\"{\i}ve particle picture and replace it by
the one based on quantum fields with their eventualizations.

The idea of eventualization may appear to be just a palliative to avoid the
contradiction between abstract theory of localization and the concrete
localization phenomena, but actually, it opens the door to quite natural
formulation of quantum physics. In fact, the notion of measurement process can
be considered as a special kind of eventualization process with amplification.
As we will discuss in a forthcoming paper \cite{OOS15}, a glossary of
\textquotedblleft quantum paradoxes\textquotedblright\ is solved by just
posing an axiom we call \textquotedblleft eventualization
principle\textquotedblright.

\begin{quote}
\textbf{Eventualization Principle:} Quantum fields can effect macroscopic
systems only through eventualization.
\end{quote}

In other words, we hypothesize that the notion of \textquotedblleft
macroscopic systems\textquotedblright---including a Schr\"{o}dinger cat--- can
be characterized, or defined, by the collection of events, formed by
perpertual eventualization.

\section*{Acknowledgments}

We would like to express our sincere gratitudes to Prof.~P. Jorgensen for
inviting us to the opportunity of contributing this paper to a Special Issue
\textquotedblleft Mathematical\quad Physics\textquotedblright\ in the Journal
\textquotedblleft Mathematics\textquotedblright. We are grateful to Prof.~S.
M. Barnett, Dr.~T. Brougham, Dr.~V. Poto\v{c}ek and Dr. M. Sonnleitner for
enlightening discussions on the occasion of one of us (H.S.) to visit Glasgow.
Similarly, we thank Prof.~M. Bo\.{z}ejko, Prof.~G. Hofer-Szab\'{o} and
Prof.~M. R\'{e}dei for encouraging discussions in Wroc\l aw and Budapest. We
are grateful to Prof.~M. Ohtsu, Prof.~M. Naruse and Prof.~T. Yatsui for their
interests in our work and instructive discussions in Tokyo. Last but not
least, we cordially thank Prof.~H. Sako and Dr.~K. Okamura for inspiring and
continuing collaboration.

\end{document}